# Direct Search for Dark Matter Particles With Very Large Detectors[*]

David B. Cline

Astrophysics Division, Department of Physics & Astronomy, University of California, Los Angeles, CA 90095 USA   dcline@physics.ucla.edu

**Summary.** We briefly discuss the expected level of supersymmetric dark matter cross-sections as a reference for dark matter detectors. We then discuss the current ZEPLIN II program as a prototype of large liquid Xenon detectors. Cryoarray is a possible cryogenic detector. Finally we discuss ZEPLIN IV and other one ton liquid Xenon detectors and the limiting backgrounds for such detectors.

## 1 Introduction

The current situation in SUSY dark matter search is this [1]:

(a)    The DAMA result is challenged by recent limits on the detection from CDMS I (2000 and 2003 results), Edelweiss and ZEPLIN I. This mixture of detectors presents a diversity of target masses and spins. A careful study is needed to see if there is any way to make these results consistent.

(b)    Nearly all calculations for the expected rate give a low value, well below current limits. At least 4 groups agree on these predictions.

(c)    The next generation detectors, Edelweiss II, CDMS II, ZEPLIN II and others, can expand the search by about a factor of $10^2$ (to $10^{-3}$ events/kg/day). At this level all detectors will likely see neutron background events.
   In order to go to a more sensitive region and possibly cover the entire SUSY expectations two developments are needed:

   (i) At least one ton discriminating detector (ZEPLIN IV, CryoArray, etc.)
   (ii)  A clear understanding of how to handle or eliminate the neutron background.
   (iii) A method to compare the different experiments.

These experiments need to get to $10^{-46}$ cm$^2$ in the detector or a neutron flux of $10^{-12}$ cm$^{-2}$ sec$^{-1}$ at the detector. We first turn to the neutron background.

## 2  Study of the Neutron Background

---

[*] Invited talk, Beyond Conference, June 2003.

Peter Smith, Hanguo Wang and the author are studying the current observations of neutrons in underground laboratories and will carry out calculations of the different sources of neutrons [2]. The required depth of the laboratory can be determined.

A schematic of our study of neutron backgrounds for the future SUSY-WIMP search is shown in Figure 1.

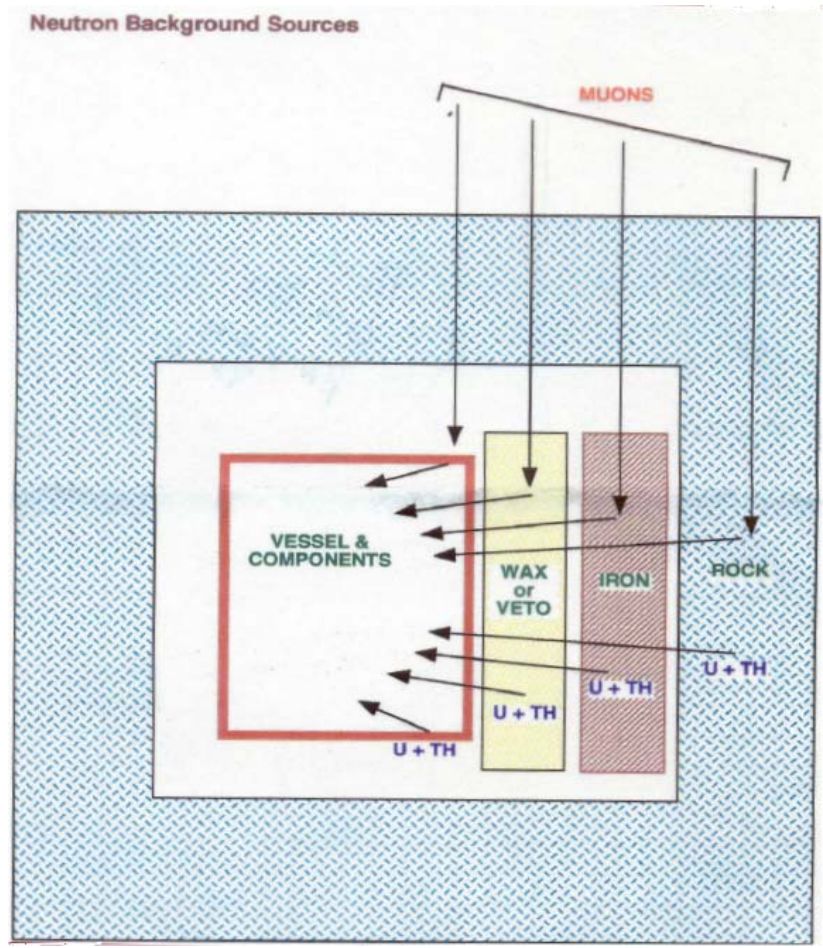

**Fig. 1.** Schematic of the neutron background at UCLA.

A recent measurement of the neutron flux at the Gran Sasso shows an unexpected feature of higher energy neutrons (see Fig. 2). For this reason a careful study of sources of neutron background is underway at UCLA [2]. Currently we do not know if a site like Boulby is deep enough to reduce the neutron background. The new SNO Lab may be deep enough in this case.

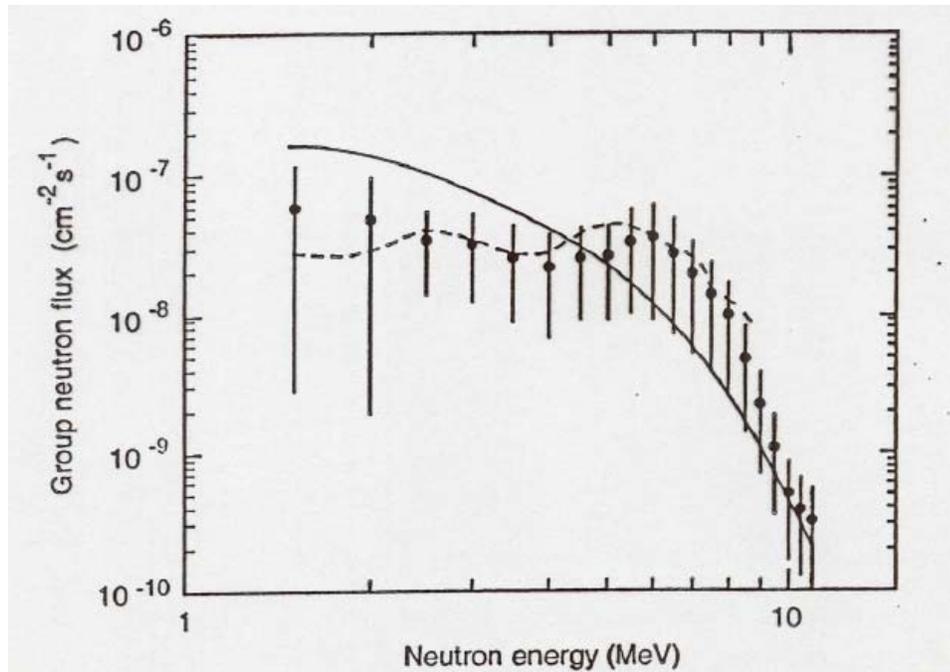

**Fig. 2.** Recent measurement of the neutron background flux at the LNGS by the ICARUS team.

The added structure in this distribution above 5 MeV is difficult to understand and perhaps could be a problem for future searches.

## 3  Sensitivity of Detectors to Different WIMP Masses and Astrophysical Uncertainties

There seems to be a great deal of confusion on how to compare the data from the DAMA, CDMS, Edelweiss and ZEPLIN I experiments. Two methods are used:
  a) Each experiment claims a certain region of exclusion or discovery in the cross-section/WIMP mass plot
  b) An attempt is made to carry out a joint analysis of the DAMA and CDMS or Edelweiss data.

In the latter case the most recent CDMS paper reports a 99.9% exclusion of the DAMA discovery region. Both kinds of analysis have problems, in my opinion.

Recently Copi and Krauss (Ref. 3) have attempted to compare the sensitivity of the DAMA and CDMS results with different types of astrophysical models of the galactic halo. They conclude that the different models do not change appreciably the ratio of DAMA to CDMS sensitivity and agree that the results are not compatible. More recently Bottino and collaborators have also examined these claims [4].

In order to compare these results more model independent methods are required. We believe the key is the minimal velocity required to produce a given energy deposition in a detector as a function of WIMP mass. This velocity is given as

$$v_{min} = \sqrt{\frac{E_t[M_w + M_A]^2}{2M_w^2 M_A}} \quad (1)$$

where $E_z$ is the detector threshold $M_w$, WIMP mass and $M_A$ target mass. We can see the importance of energy threshold and target mass in the two limiting cases:

$$M_w \ll M_A$$
$$v_{min} \approx \sqrt{\frac{E_t M_A}{2M_w^2}} \quad (2)$$

$$M_w \gg M_A$$
$$v_{min} \approx \sqrt{\frac{E_t}{2M_A}} \quad (3)$$

for $M_w = M_A$

$$v_{min} = \sqrt{\frac{2E_t}{M_w}} \quad (4)$$

At present the greatest region of uncertainty is for small $M_w$. As $M_w$ decreases the $v_{min}$ increases making it more difficult for all experiments to reach low $M_w$. However the lower threshold CDMS detector would seem to have the advantage. Most believe this region has already been mainly excluded by LEP data and theory.

In Figure 3 we show different possible halo velocity distributions. I also plot the $v_{min}$ for the different detector. While a decrease in the mean velocity would seem to indicate an enhanced annual variation signal, in DAMA the required $v_{min}$ pushes the velocity up, especially at low WIMP mass. Therefore these effects would seem to cancel. This is why the analyses of Copi and Krauss show a near independence of the halo models, I believe (see Reference 3).

In a very simpleminded approach the larger the $v_{min}$ the smaller would be the annual variation signal in general. This would seem to disfavor the study of annual variation signals for low mass WIMPs. In addition the larger the $v_{min}$ the less likely that a low velocity dispersion galactic halo model would be favored for any annual variation study. See Reference [5] and Reference [6].

These factors need to be considered for future very large detectors. The lowest possible threshold should be obtained. In the case of Liquid Xenon there is the possibility to employ the signal as was shown by the UCLA/Torino group [Reference 7].

# 4  Dark Matter Search for Current Detectors

The current status of the search for WIMPs is somewhat confused; a recent summary can be found in Ref. 1:
1. The sensitivity level is about 1/5 event/kg/day [1];
2. The DAMA group is making claims for a discovery; by observation of a possible annual signal variation [1];
3. The CDMS group has shown that their data and those of DAMA are incompatible to 99.5% confidence level against the observation of WIMPs [1].
4. The Edelweiss and ZEPLIN I groups have excluded the mean value of the DAMA results [6].
5. Recent ZEPLIN I results give the best current limits, well below the DAMA results [private communication, N. Smith].

Most current and next generation detectors are listed in Table 1 [1].

## Leading Searches for Dark Matter

| Project | Location | Start Date | Primary Detector Type | Primary Detector Material | Primary Detector Mass (kg) | Discrimination Detector Types(s) |
|---|---|---|---|---|---|---|
| UKDMC | Boulby, UK | 1997 | Scintillation | Sodium iodide | 5 | None |
| DAMA | Gran Sasso, Italy | 1998 | Scintillation | Sodium iodide | 100 | None |
| ROSEBUD | Canfranc, Spain | 1999 | Cryogenic | Aluminum oxide | 0.05 | Thermal |
| PICASSO | Sudbury, Canada | 2000 | Liquid droplets | Freon | 0.001 | None |
| SIMPLE | Ristrel, France | 2001 | Liquid droplets | Freon | 0.001 | None |
| DRIFT | Boulby, UK | 2001 | Ionization | Carbon disulfide gas | 0.16 | Directional |
| Edelweiss | Frejus, France | 2001 | Cryogenic | Germanium | 1.3 | Ionization, thermal |
| ZEPLIN I | Boulby, UK | 2001 | Scintillation | Liquid Xenon | 4 | Timing |
| CDMS II | Soudan, Minn., US | 2003 | Cryogenic | Silicon, germanium | 7 | Ionization, thermal |
| ZEPLIN II | Boulby, UK | 2003 | Scintillation | Liquid Xenon | 30 | Ionization, scintillation |
| CRESST II | Gran Sasso, Italy | 2004 | Cryogenic | Calcium tungsten oxide | 10 | Scintillation, thermal |
| GENIUS-TF | Gran Sasso | 2003 | Ionization | Germanium | 10kg | Ionization |
| GENIUS | Gran Sasso | | Ionization | Germanium | 100kg | Ionization |

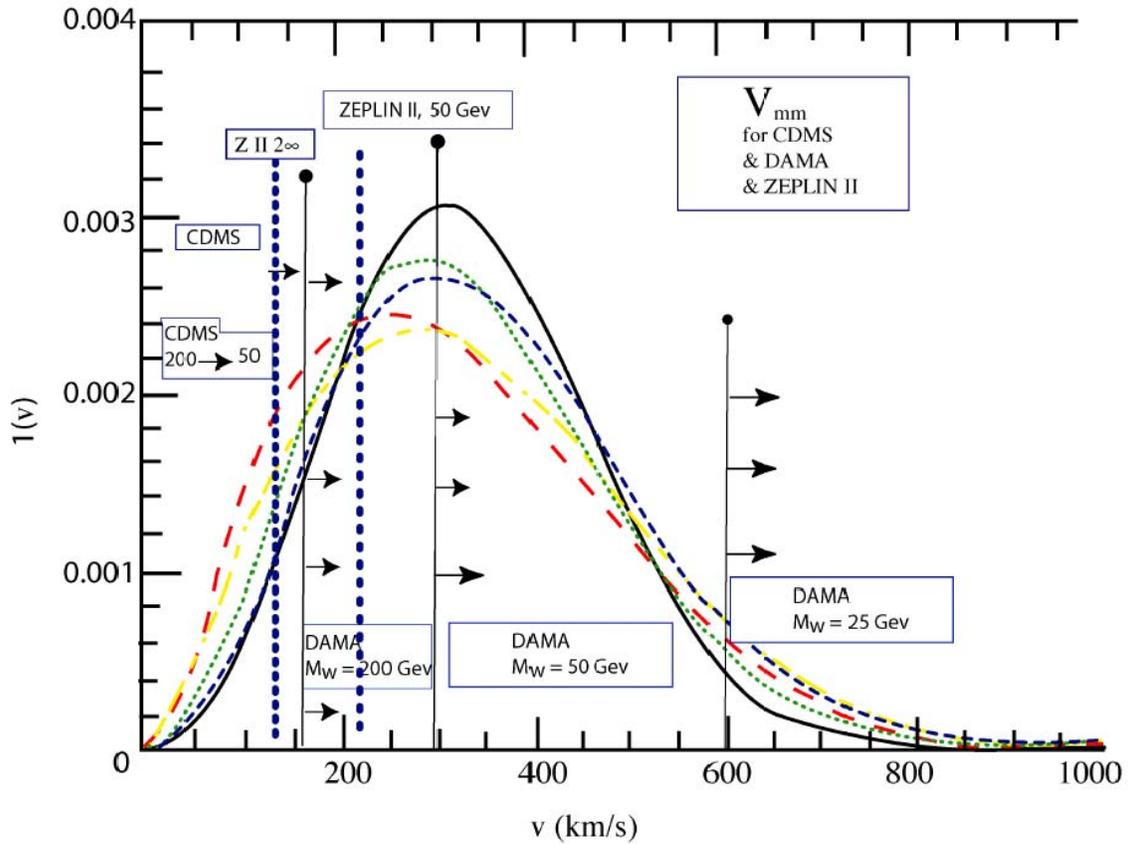

**Fig. 3**. Schematic of the halo velocity distribution with minimal velocities for CDMS, DAMA and ZEPLIN II; the figure is modified from A. Green, P.R.D. **68**, 023004 (2003)(Reference 5).

## 5 Next Generation of Detectors

We expect ZEPLIN II, Edelweiss II and CDMS II to be some of the key detectors that are now under construction. Liquid Xenon is of great importance [7]. The key properties of Xenon are given in Table2.

Starting in the early 1990s, the UCLA/Torino ICARUS group initiated the study of liquid Xe as a WIMP detector with powerful discrimination. The basic mechanism of detection in Liquid Xenon is shown in Fig. 4. WIMP interactions are clearly discriminated from all important background by the amount of free electrons that are drifted out of the detector into the gas phase where amplification occurs. In Fig. 5, we show the resulting separation between backgrounds and simulated WIMP interactions (by neutron interaction). It is obvious from this plot that the discrimination is very powerful [7].

Construction has begun on a large two-phase detector to search for WIMPs. The UCLA/Torino group has formed collaboration with the UK Dark Matter team to construct a 40-kg detector (ZEPLIN II) for the Boulby Mine underground laboratory (Fig. 6).

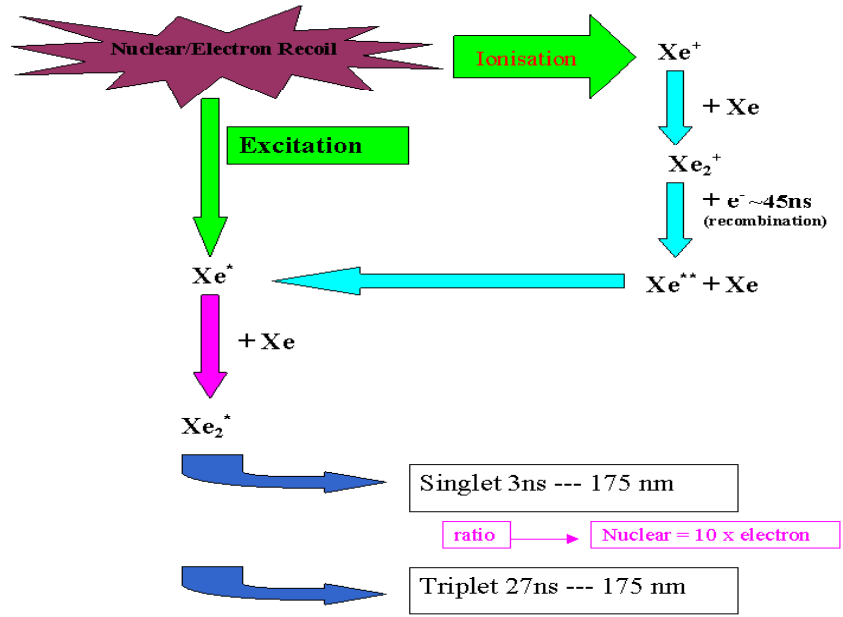

**Fig. 4.** The light generation process in Liquid Xenon.

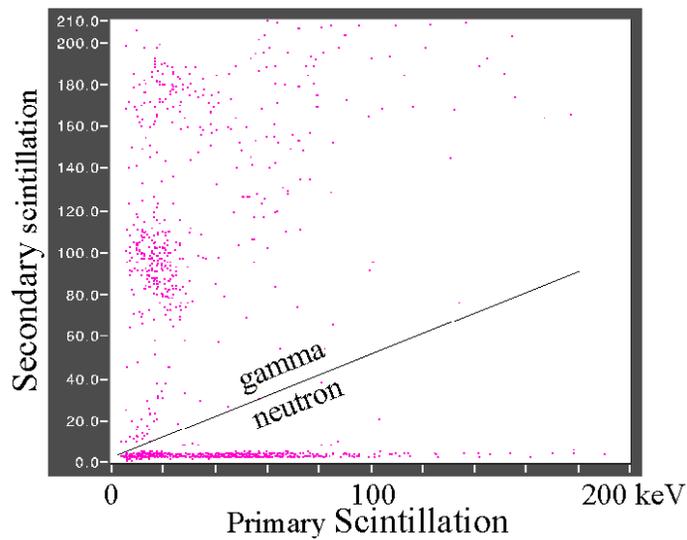

**Fig. 5.** Discrimination results for Liquid Xenon (2 phases) [7].

**TABLE 2.**

**Liquid Xenon as a WIMP Detector**

1. Large mass available - up to tons.

- Atomic mass: 131.29
- Density: 3.057 gm/cm$^3$
- $W_i$ value (eV/pair) 15.6 eV
- No long-lived isotopes of xenon

2. Drift velocity: 1.7 mm/μ.s @ 250V/cm field
    X Scintillation wave length: 175 nm

- Decay time: 2 ns → 27 ns

3. Light yield > NaI, but intrinsic scintillator (no doping)

⇒ Excimer process very well understood

⇒ First excimer laser was liquid xenon in 1970!

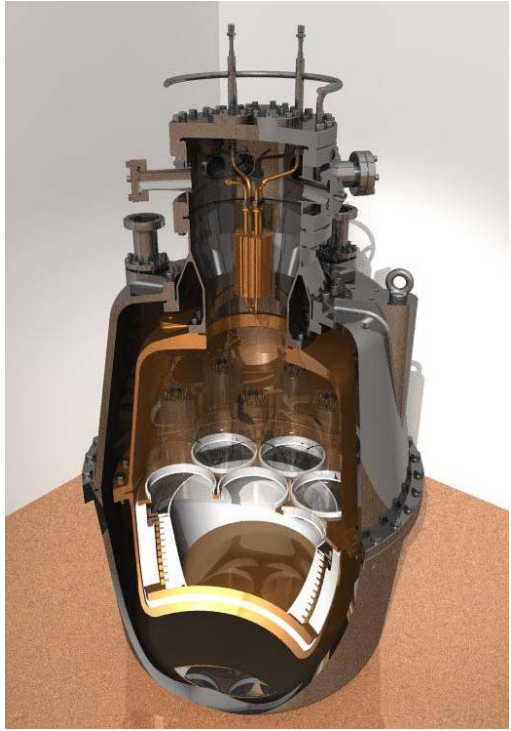

**Fig. 6.** ZEPLIN II detector for Boulby Labs.

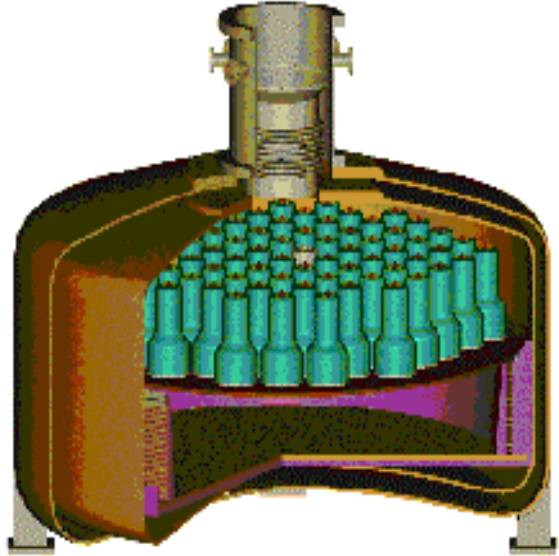

**Fig. 7.** Schematic of a one ton detector (ZEPLIN IV) scaled up from ZEPLIN II.

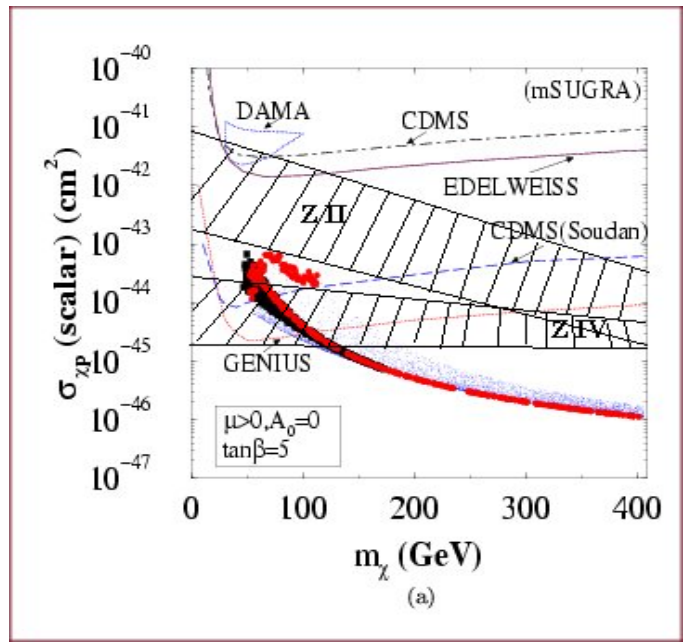

**Fig. 8.** Expected cross-sections for SUSY WIMPs in SUGRA models (P. Nath) and some recent search results. The range of sensitivity for ZEPLIN II and IV are shown on the modified plot [2].

## 6 One Ton Detector

Liquid Xenon is the most likely material to be used for very large detectors due to the properties of the Xenon (Table 2) and the excellent discrimination possibility as demonstrated by the UCLA/Torino group. The basic discrimination is shown in Figure 5.

Continuation of the R&D effort with Liquid Xenon attempts to amplify the very weak signal. The second idea to test is inserting a CsI internal photo cathode to convert

UV photons to electrons that are subsequently amplified by the gas phase of the detector. This detector is being assembled at Rutherford Laboratory now for installation in the Boulby underground laboratory later this year.

A simple scaled-up version of ZEPLIN II (ZEPLIN IV) to a one ton detector is shown in Figure 7. The reach of such a detector will be limited by backgrounds most likely [1]. In Figure 8 we show the possible sensitivity of the one ton detector. Most of the SUSY-WIMP preferred region is covered. To cover the lowest region may require a very deep site for the detector to reduce neutron background.

I wish to thank H. Wang, Y. Seo and other members of the UCLA Dark Matter team as well as members of the UKDMC for advice.